\documentclass[aps,pra,twocolumn,superscriptaddress,floatfix,
nofootinbib,showpacs,longbibliography]{revtex4-1}

\usepackage{flushend}
\usepackage{xcolor}
\usepackage{cancel}
\usepackage[utf8]{inputenc}  
\usepackage[T1]{fontenc}     
\usepackage[british]{babel}  
\usepackage[sc,osf]{mathpazo}
\usepackage[scaled=0.86]{berasans}  
\usepackage[colorlinks=true, citecolor=blue, urlcolor=blue]{hyperref}  
\usepackage{graphicx} 
\usepackage[babel]{microtype}  
\usepackage{amsmath,amssymb,amsthm,bm,amsfonts,mathrsfs,bbm} 

\usepackage{xspace}  
\definecolor{wrwrwr}{rgb}{0.3803921568627451,0.3803921568627451,0.3803921568627451}
\definecolor{rvwvcq}{rgb}{0.08235294117647059,0.396078431372549,0.7529411764705882}
\usepackage{pgf,tikz,pgfplots}
\usetikzlibrary{calc}
\usepackage{mathrsfs}
\usetikzlibrary{arrows,snakes}
\usetikzlibrary{positioning}
\usepackage{xcolor}
\usepackage{appendix}
\usepackage{multirow}
\usepackage{array}
\usepackage{bigstrut}
\usepackage{braket}
\usepackage{color}
\usepackage{natbib}
\usepackage{multirow}
\usepackage{float}
\usepackage[caption = false]{subfig}
\usepackage{xcolor,colortbl}
\usepackage{color}

\newcommand{\be}{\begin{equation}}
\newcommand{\ee}{\end{equation}}
\newcommand{\ba}{\begin{eqnarray}}
\newcommand{\ea}{\end{eqnarray}}

\newcommand{\tr}{\operatorname{Tr}}

\definecolor{rvwvcq}{rgb}{0,0,1}
\usepackage{tabu}

\def\>{\rangle}
\def\<{\langle}

\makeatletter
\DeclareRobustCommand{\cev}[1]{%
	{\mathpalette\do@cev{#1}}%
}
\newcommand{\do@cev}[2]{%
	\vbox{\offinterlineskip
		\sbox\z@{$\m@th#1 x$}%
		\ialign{##\cr
			\hidewidth\reflectbox{$\m@th#1\vec{}\mkern4mu$}\hidewidth\cr
			\noalign{\kern-\ht\z@}
			$\m@th#1#2$\cr
		}%
	}%
}
\makeatother

\begin{document}

\title{Independence of work and entropy for equal
	energetic finite quantum systems: Passive state
	energy as an entanglement quantifier}

\author{Mir Alimuddin}
\email{aliphy80@gmail.com}
\author{Tamal Guha}
\author{Preeti Parashar}    
\affiliation{Physics and Applied Mathematics Unit, Indian Statistical Institute, 203 B T Road, Kolkata-700108, India.}


\begin{abstract} 
Although entropy is a necessary and sufficient quantity to characterize the order of work content for equal energetic (EE) states in the asymptotic limit, for the finite quantum systems, the relation is not so linear and requires detail investigation. Toward this, we have considered a resource theoretic framework taking the energy preserving operations (EPO) as free, to compare the amount of extractable work from two different quantum states. Under EPO, majorization becomes a necessary criterion for state transformation. It is also shown that the passive state energy is a concave function and for EE states it becomes proportional to the ergotropy in absolute sense. Invariance of the passive state energy under unitary action on the given state makes it an entanglement measure for the pure bipartite states. Further, due to the non additivity of passive state energy for the different system Hamiltonians, one can generate Vidal$'$s monotones which would give the optimal probability for pure entangled state transformation. This measure also quantifies the ergotropic gap which is employed to distinguish some specific classes of three-qubit pure entangled states. 
\end{abstract}
\maketitle
\section{Introduction}
Work is the most fundamental observable in standard thermodynamics. The {\it'second law of thermodynamics'} rules how much maximum work one is able to extract from a thermodynamic system in a cyclic process. Generally, work is quantified by the free energy $F=U-TS$, where $U,S,T$ are respectively the internal energy, entropy and the corresponding bath temperature \cite{zemansky1998heat}. Comparing states on the basis of work extraction through a bath assisted thermodynamical process is completely equivalent with the free energy comparison and this feature is also exhibited by quantum systems in the asymptotic limit \cite{PhysRevLett.111.250404, skrzypczyk2014work}. Considering work extraction from closed systems, if we make this comparison only between equal energetic $(EE)$ states, then it solely depends on their entropy. The higher entropic states have less work content and vice versa\cite{Allahverdyan_2004,PhysRevE.87.042123}. Axiomatic thermodynamics says that $A \rightarrow B$ transformation is possible adiabatically, if and only if $S(A) \leq S(B)$ \cite{LIEB19991}. The thermodynamical process also induces an order in the extractable work i.e., $W(A) \geq W(B)$. But for the finite quantum systems, since all the axioms are not satisfied there exist states which are not inter convertible adiabatically\cite{lieb2013entropy, lieb2014entropy, PhysRevLett.117.260601}. The same problem has been encountered independently in several other resource theoretic frameworks \cite{PhysRevA.67.062104, horodecki2013fundamental, Brandao3275,ng2018resource,PhysRevLett.89.037903}. As an example there exist bi-partite pure entangled states which are not convertible from each other under Local operation and classical communication $(LOCC)$ \cite{PhysRevLett.83.436,PhysRevLett.83.1046}. Moreover, in the asymptotic limit of entanglement theory this convertibility directed by the marginal entropy of the given states\cite{PhysRevA.53.2046,PhysRevLett.89.037903}, whereas in the finite copy, convertibility would hold only when there exists majorization criterion among the marginals. This leads to the existence of many independent monotones simultaneously, that converge to the entropy in the asymptotic limit.
\par
In this article, we have studied the similar kind of finite copy inconsistency in thermodynamic domain along with the characterization of a class of states, for which the inconsistency can be removed. Precisely, we have introduced an operation on the finite particle regime, namely {\it energy preserving operation} (EPO) which is very similar to the isothermal process in the thermodynamic limit. Alike the role of entropy under this isothermal process to direct the state transformation, majorization is an indicator for EPO. More formally, we have shown that if two same energetic state $(A\text{ and }B)$ are one way convertible under EPO $(A\to B)$, then $A$ is majorized by $B$ necessarily. On the other hand, if the conversion between two EE state is not possible under EPO in either direction, then the order of extractable work for these two states has no relation with their entropy. Here, we consider resource theoretic approach where the majorization becomes the sufficient criterion for state convertibility under any EPO. The equivalence relation between the order of entropy and the order of work exists for states which obey majorization.
 In this article we have shown that entropy and ergotropy (maximum extractable work under unitary operation \cite{Allahverdyan_2004}) become independent monotones under the $EPO$. This implies that entropy is neither necessary nor sufficient condition for extractable work from a closed system. Single shot work extraction\cite{aaberg2013truly,horodecki2013fundamental} from a state $\rho$ under a fixed temperature bath is defined as $W_S = \frac{1}{\beta}D_{min}(\rho_{\omega}\parallel\tau_{\beta})$ ($\rho_{\omega}$ is dephased $\rho$ in energy basis), which is another independent monotone.

\par

For a given state $\rho$, the lowest energetic state with the same spectrum is called passive state, represented by $\rho^p$. We have shown that just like the entropy, passive state energy also is a concave function. For the $EE$ states, ergotropy and passive state energy become proportional (in the absolute sense). On the other hand, any concave function which remains invariant under unitary can be a monotone under $LOCC$ \cite{vidal2000entanglement}. Entropy is a good measure of entanglement for pure bipartite states since it gives the rate of distillation under $LOCC$ in the asymptotic limit. Due to its additivity property it is not able to provide optimal probability of state transformation for the finite copies. However, very recently quantum entanglement has been investigated using non additive entropies \cite{PhysRevA.100.032327,san2018hamming}. In this paper, we have established the passive state energy of the marginals as an entanglement monotone which is by nature a non-additive quantity.  It quantifies the collaborative advantage in work extraction: suppose Alice and Bob shared a pure entangled state where they can act a suitable unitary jointly or locally to extract maximum work. It is obvious that one can get more work due to the joint unitary action. This extra gain is the collaborative advantage and is called ergotropic gap\cite{PhysRevA.99.052320, PhysRevE.93.052140}. From this perspective, we can call passive state energy as a {\it thermodynamic measure of entanglement}. In addition, it generates Vidal's entanglement monotones \cite{PhysRevLett.83.1046} under different system Hamiltonians and gives the optimal rate in finite copy state transformation due to it's non additive nature. A convex hull representation is given for the mixed entangled states which makes it a faithful measure and eventually non-monogamous\cite{PhysRevLett.117.060501}. 

\par

In our study, we have shown that passive energy is equal with the ergotropic gap up to some proportional factor, so we can consider it as an equivalent measure of entanglement. For $2\times2\times2$ systems we have shown that entanglement between any cut $A|BC$ is equal to the sum of the ergotropic gap of $AB$ and $AC$ and can be extended for higher dimension to give an upper bound on the entanglement in a particular partition. Furthermore, unlike $GHZ$, for $W$ class of states help from the third party $(C)$ is not necessary to get collaborative advantage and entanglement of $A|BC$ is captured by $A|B$ only. We further use this thermodynamic measure to classify some tripartite pure entangled states.
\par

Through out this article, the spectrum of any given state has been taken in non-increasing order, whereas energy eigenvalues for the Hamiltonian are taken in non-decreasing order.

\section{Framework}
\subsection{Work Extraction from a closed system}
\begin{figure}[htb!]
    \centering
    \includegraphics[height=5.5cm,width=9cm]{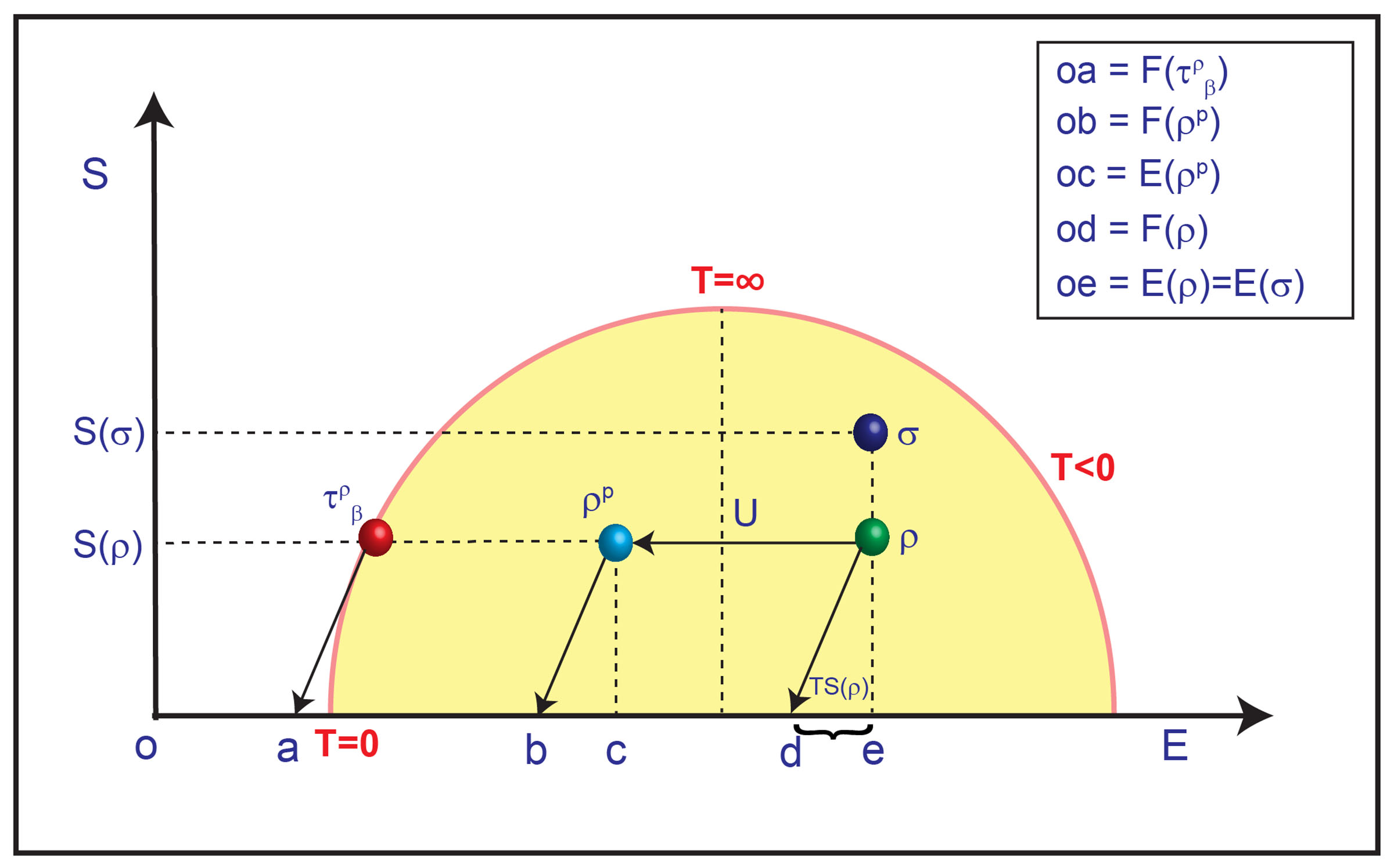}
    \caption{\label{figure1} Maximum extractable work from a finite closed system  $\rho$ is given by $W_e(\rho)= E(\rho)-E(\rho^p)=F(\rho)-F(\rho^p)$ which is bounded by the $W_{th}(\rho)=F(\rho)-F(\tau^{\rho}_{\beta})$. In the asymptotic limit, for the $EE$ states $\rho$ and $\sigma$, lower entropy is always a better resource for work extraction but for the finite copy if the majorization condition holds then only we can make the above claim.}
   \end{figure}

 For a finite closed quantum system, the authors in \cite{Allahverdyan_2004} have introduced an equivalent scheme of standard adiabatic process, where the system is governed by arbitrary unitaries, which implies that not only entropy but rather the whole spectrum remains unchanged.  
 Under this process the maximum amount of extractable work is called ergotropy defined by $W_e(\rho) = Tr(\rho H_S)-\min_{U(t)}\{U(t)\rho U^{\dagger}(t)H_S\} = Tr(\rho H_S)-Tr(\rho^p H_S)$, where the governing  Hamiltonian is $H_S=\sum\limits_{i=0}^{d-1}\epsilon_{i}|\epsilon_i\rangle\langle \epsilon_i|$. Also $\rho^p = \sum\limits_{i=0}^{d}p_i|\epsilon_i \rangle \langle \epsilon_i |$, where $p_i \geq p_{i+1}$ if and only if $\epsilon_i \leq \epsilon_{i+1}$ \cite{pusz1978passive,lenard1978thermodynamical,PhysRevE.91.052133} is the passive state corresponding to $\rho = \sum\limits_{i=0}^{d}p_i |i\rangle \langle i |$. From the definition, there always exist an equal entropic but lowest energetic state called Gibb's state, defined by $\tau_{\beta} = \frac{e^{-\beta H_S}}{Z}$,
  (where $Z=\sum_{i} exp(-\beta \epsilon_i)$ is the partition function) which may not be always achievable from the initial state $\rho$  by a unitary action since their spectrum are not same. In the asymptotic limit it has been shown that $\rho^{\otimes n} \rightarrow \tau^{\otimes n}_{\beta}$ $(n \rightarrow \infty)$ transformation is possible unitarily and maximum extractable work on an average becomes equal to the thermodynamic work $W_{th} = Tr(\rho H_S)-Tr(\tau_{\beta} H_S) = E(\rho)-TS(\rho)+TlnZ$. This sets an upper limit on the ergotropic work $W_e$ \cite{PhysRevE.87.042123}. The whole picture can be visualized by the simple energy-entropy diagram \ref{figure1} \cite{bera2018thermodynamics, bera2019thermodynamics}.  
 
 \par
 
 Below we make a comparison of maximum extractable work {\it(ergotropy)} from two closed systems on the basis of their given information. 
 
 \par
 
  {\it (i) States with unequal internal energy and entropy:} If
  $\rho \succ \sigma $ and $Tr(\rho H_s) > Tr(\sigma H_s)$ then it implies $W_e(\rho) > W_e(\sigma)$, otherwise we need to have the complete information \cite{ci} about the states in order to make an appropriate comment about the maximum extractable work. \\

  {\it (ii) States with unequal internal energy but same entropy:} According to standard thermodynamics, equal entropic higher energetic states always have more free energy and so give more work. But in the finite limit, closed systems change unitarily and comparison is not possible unless they have the same spectrum. In such special cases, transformation from higher to lower energetic state gives positive work.\\

  {\it (iii) States having equal internal energy but different entropy:} Although in the thermodynamic limit, lower entropy provides higher work but in the finite quantum systems entropy cannot tell us anything about the ergotropy. A sufficient criterion for ergotropy, i.e; if $\rho \succ \sigma$ then $W_e(\rho) \geq W_e(\sigma)$ has been given in \cite{Allahverdyan_2004}.\\
  
  \par 
  
   
   \subsection{Work extraction in the presence of bath:}
\begin{figure}[htb!]
              \centering
              \includegraphics[height=4.5cm,width=7cm]{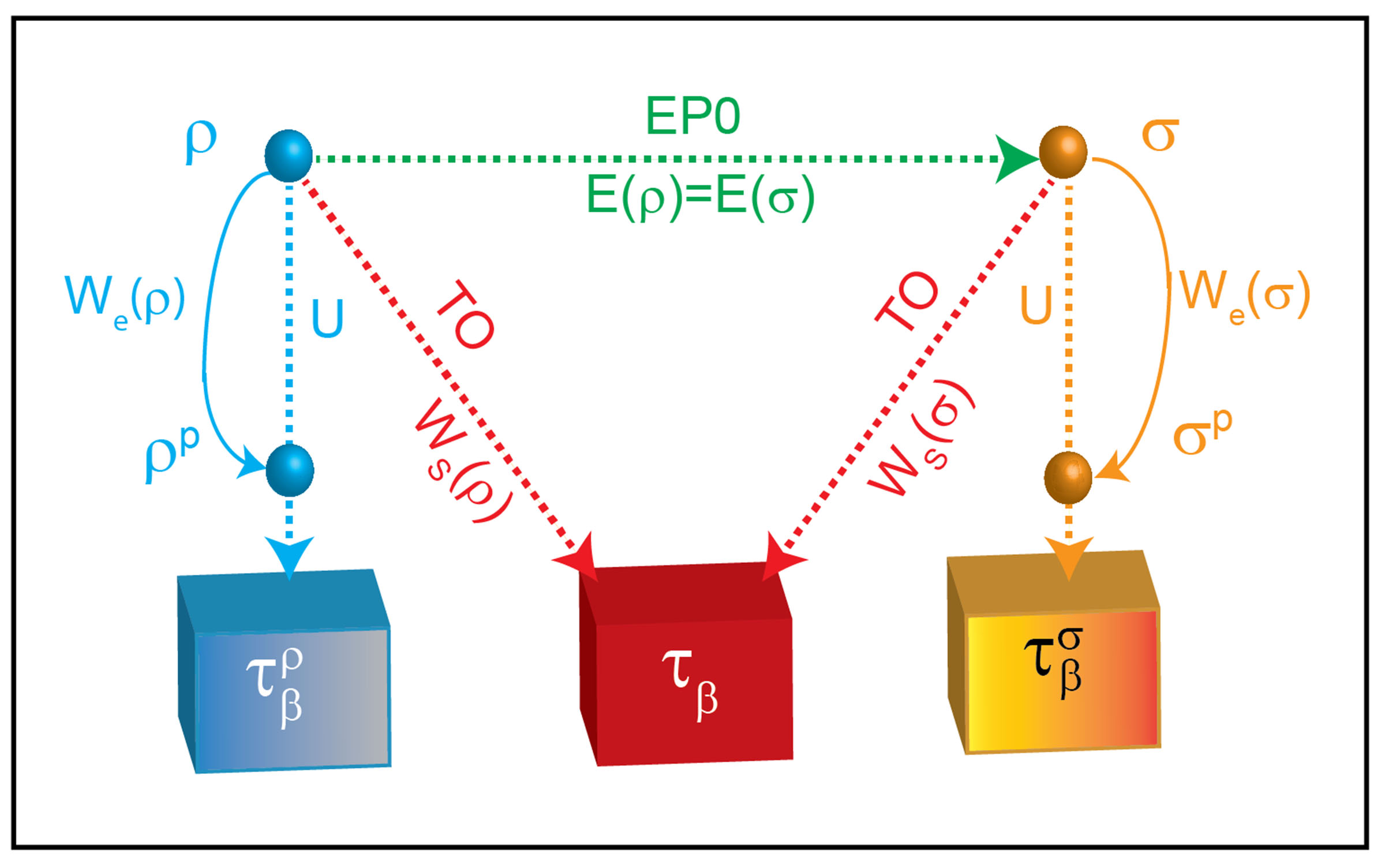}
              \caption{\label{figure2}{\it Independence of work and entropy on the finite $EE$ states:} Let $\rho \rightarrow \sigma$ under $EPO$  (i.e., $\rho \succ \sigma$). Then $S(\rho) \leq S(\sigma)$ and the order on the maximum extractable work under unitary as well as thermal operations is $W_e(\rho)\geq W_e(\sigma)$ and $W_S(\rho) \geq W_S(\sigma)$ respectively, like in the asymptotic limit. But if the two finite $EE$ states are not connected by $EPO$ then the extractable work and entropy becomes completely independent since entropy($S$), ergotropy $(W_e)$, single shot extractable work $(W_S)$ all are independent monotones under this operation. }
             \end{figure}

        Work extraction from a finite quantum system in the presence of a bath has been discussed in statistical as well as resource theoretical approach \cite{skrzypczyk2014work, horodecki2013fundamental, Brandao3275,aaberg2013truly}. Although in the asymptotic limit it becomes consistent with the classical results and is quantified by the free energy\cite{PhysRevLett.111.250404}, in the finite limit it becomes less and $0$-free energy($\alpha=0$) is the quantifier of single shot work extraction. Let us consider a state $\rho$ with the corresponding bath of inverse temperature $\beta=\frac{1}{kT}$. Bath particles having the Gibbsian form called thermal states are defined by $\tau_{\beta} = \frac{e^{-\beta H_B}}{Z}$,
        with the governing Hamiltonian $H_B=\sum\limits_i\epsilon_{i}|\epsilon_{i}\rangle\langle \epsilon_{i}|$ and the corresponding partition function $Z=\sum_{i} exp(-\beta \epsilon_i)$.
        All other states are considered as athermal and in the asymptotic limit a more resourceful (in terms of free energy) to less resourceful state transformation is possible by the thermal operation. In the finite limit, infinitely many free energies $F_{\alpha}$ for $0\leq \alpha \leq \infty$ are necessary to direct state transformation, and among them $W_S(\rho_{\omega})=F_0(\rho)-F_0(\tau_{\beta})= \frac{1}{\beta}D_{0}(\rho_{\omega}\parallel\tau_{\beta})$ (depicted in the figure \ref{figure2}) quantifies the maximum extractable work deterministically (single shot work extraction), where $\rho_{\omega}= \sum_{E}\Pi_{E}\rho\Pi_{E}$  is the dephased $\rho$ in energy eigen basis, $D_{0}(\rho_{\omega}\parallel\tau_{\beta}):= -\log \tr({\Pi_{\rho_{\omega}}}\tau_{\beta})$ where $\Pi_{\rho_{\omega}}$ is the the projector onto the support of $\rho_{\omega}$ . 
        
        \subsection{Energy Preserving Operation (EPO)}
        Recently energy preserving paradigm has been introduced in \cite{PhysRevA.96.022327}, in the context of zero energy cost quantum operations, which we briefly discuss. Since the main aim is to preserve the energy of the system, we are not bound to the energy conserving unitaries only. Rather the system can interact with the environment through some interaction Hamiltonian $H_{int}(t)$ for $t_1 \leq t \leq t_2$ such that through out the process, energy remains constant. The system and the environment (governed by $H_{E}$) together evolve as a closed system under the corresponding unitary
        \begin{equation}
        U(t) = \overrightarrow{exp}\{-\frac{i}{\hbar} \int_{t_1}^{t_2} [H_S + H_E + H_{int}(t)]dt\},
        \end{equation}
        
        Since we want to implement this process in the thermodynamic paradigm, energy of the closed joint system should be conserved i.e;
        \begin{equation}\label{total}
        [U, H_S+H_E]=0,
        \end{equation} 
         which in turn implies, 
         \begin{equation}
         [ H_{int},H_S + H_E]=0.
         \end{equation}
         It means that although some external field is applied for making interaction between the system and the environment, total work done by the field is always zero.
         
         \par
         
         In our scenario $[U,H_S]=0$ since the energy of the system has to be conserved. As a consequence of the joint (Eq.\ref{total}) as well as system energy conservation, the energy of the environment remains unchanged i.e., $[U,H_E]=0$. Let the system and environment be initially uncorrelated which under the energy conserving map evolves to
         \begin{equation}
         \Lambda (\rho_S) = Tr_E[U(\rho_S \otimes \rho_E)U^{\dagger}]= \sum\limits_{k}^{} M_k \rho_S M^{\dagger}_k.
         \end{equation}
         
          The Kraus operators $M_k$'s have been shown \cite{PhysRevA.96.022327} to commute with  $H_S$ i.e;
         \begin{equation}
         [M_k, H_S]=0,
         \end{equation}
         which implies that they should be diagonal or Bloch diagonal in the energy basis.      
        
        \subsection{Majorization relation}
        {\it Definition:} A state $\rho$ majorizes a state $\sigma$ i.e. $\rho \succ \sigma$ if,
         \begin{equation}
         \sum\limits_{i=1}^{k}p_{i}^{\downarrow} \geq \sum\limits_{i=1}^{k}q_{i}^{\downarrow} ~~~~ (1 \leq k \leq n-1)
         \end{equation}
          and
         \begin{equation}
          \sum\limits_{i=1}^{n}p_{i}^{\downarrow} = \sum\limits_{i=1}^{n}q_{i}^{\downarrow},
         \end{equation}
         where $\{p_{i}^{\downarrow}\}^{n}_{i=1} \in \mathbb{R}^{n} $ and $\{q_{i}^{\downarrow}\}^{n}_{i=1} \in \mathbb{R}^{n}$ are the spectrum of $\rho$ and $\sigma$ respectively, arranged in non-increasing order.
         In case of different dimension some extra zeros are added for completion of the above condition. Majorization criterion has great implication in state transformation in various resource theories \cite{PhysRevA.67.062104,ng2018resource,PhysRevLett.83.436}. If $\rho \succ \sigma$ then it implies that $S(\rho) \leq S(\sigma)$ (but not the reverse) and  $\rho \rightarrow \sigma$ transition is possible under noisy evolution\cite{PhysRevA.67.062104}.
         
         \section{Resource Theory of Ergotropy}
         
            In the basic structure of a resource theory, one typically considers free states and free operations. A state is called free if it cannot be transformed to a resource via any free operation of the theory. The free operations are those which cannot produce a more resourceful state from a given state. This also means that it takes free state to a free state only. In our resource theory, the class of free operations are nothing but the EPO. Since they are incoherent operations in energy eigenbasis, they would always map a diagonal state (in energy eigenbasis) to a diagonal state. So the first condition of a free state here is that it must be diagonal in energy eigenbasis. Given a quantum system with the Hilbert space $\mathcal{H}_S$ and Hamiltonian $H_S$, the allowed free operations are the completely positive and trace preserving (CPTP) maps $\Lambda_{EPO}: \mathcal{D(H_S)} \rightarrow \mathcal{D(H_S)}$ of the form
         
         \begin{equation}
         \Lambda_{EPO}(\rho_S)=Tr_E(U_{SE}(\rho_S \otimes \tau_E^{\beta})U^{\dagger}_{SE}),
         \end{equation} 
         
         where, $\tau_E^{\beta}=e^{-\beta H_E}/tr(e^{-\beta H_E})$, $\beta := {1}/{kT}, -\infty \leq \beta \leq \infty$. The $\tau^{\beta}_S$'s are the free states because it has been proved in Theorem (\ref{theorem}) that $\Lambda_{EPO}(\tau^{\beta}_S)=\tau^{\beta}_S$ where $\tau^{\beta}_S$'s are the constant energetic maximum entropic states associated with $H_S$. In the special case of a non-degenerate two-level system an inverse temperature $(\beta)$ can be assigned to any density matrix commuting with the Hamiltonian. In the case of EPO map the energy of the closed system $(SE)$, system$(S)$, and environment$(E)$ are conserved which implies $[U_{SE},H_S+H_E]=0, [U_{SE},H_S]=0$ and $[U_{SE},H_E]=0$. The states with $\beta \geq 0$ are the Gibb's states (equilibrium) for which temperature $(T)$ is defined as a property of the system whereas $\beta < 0$ denotes  population inversion states. Similar kind of description has been given in \cite{PhysRevLett.111.250404,janzing2000thermodynamic}. Apart from global energy preservation, the system's energy also needs to be conserved which makes EPO a subclass of thermal operations $(TO)$. Under $TO$, free energy of the system always decreases and the Gibbs state is free for being lowest free-energetic state at a given temperature $T$. So free energy is also a monotone under EPO since $EPO \subseteq TO$. As the $EPO$ keeps the energy constant, monotonicity of free energy implies monotonicity of entropy. So under our resource theory, the space of free states has been extended from Gibb's states $(0 \leq \beta \leq \infty)$ to all $\tau^{\beta}(-\infty \leq \beta \leq \infty)$ since they are the lowest free-energetic (highest entropic) states.

             \par
             
             By definition EPO is a unital map since it preserves identity operator i.e.
             \begin{equation*}
             \Lambda_{EPO}(\frac{I}{d}) = \frac{I}{d}.
             \end{equation*}
             
         The state transformation under the unital map can be necessarily and sufficiently characterized by the majorization criterion \cite{chiribella2017microcanonical} as, 
         \begin{equation}
         \rho \longrightarrow \sigma ~~~~~~iff~~~~~~ \rho \succ \sigma.
         \end{equation}  
         
         So if $E(\rho)=E(\sigma)$ and there exists $\Lambda_{EPO}$, such that $\Lambda_{EPO}(\rho)=\sigma$, then $\rho\succ\sigma$. However, it is not true in the other way round, i.e., if $\rho\succ\sigma$ and $E(\rho)=E(\sigma)$ it does not necessarily guarantee the existence of an EPO, such that $\Lambda_{EPO}(\rho)=\sigma$.  
          
          It is also known that $\rho \succ \sigma$ implies $S_{\alpha}(\rho) \leq S_{\alpha}(\sigma)$, where $S_{\alpha}(\rho)=\frac{1}{1-\alpha}\log Tr(\rho^{\alpha})$ for $\alpha \in (0,1) \cup (1,\infty)$ are the Renyi entropies. For $\alpha=1$ it becomes Von Neumann entropy $S(\rho) = \lim_{\alpha\to1} S_{\alpha}(\rho)$ and other two extreme entropies are $S_0 = \lim_{\alpha\to0} S_{\alpha}(\rho)= \log rank(\rho)$ and $S_{\infty} = \lim_{\alpha\to\infty} S_{\alpha}(\rho)=-\log \lambda_{\max}$, where $\lambda_{max}$ is the maximum eigenvalue for the corresponding state. Under $EPO$ map all Renyi entropies are monotones.

         Searching for other monotones, it turns out that the energy of the corresponding passive states also increases under $EPO$ (shown in \ref{passivemonotone})
         \begin{equation}\label{monopas}
         \rho \succ \sigma~~~ \Longrightarrow~~~ E(\rho^p) \leq E(\sigma^p)
         \end{equation}
       Since $E(\rho)=E(\sigma)$, Eq.(\ref{monopas}) implies that $W_e(\rho) \geq W_e(\sigma)$. 
          So ergotropy turns out to be an independent (of entropy) monotone under the free operation $EPO$. When the states cannot be transformed into each other under $EPO$ or are incomparable under majorization, then entropy is unable to provide the hierarchy on ergotropy unlike in the asymptotic limit, i.e., there exist higher entropic states having higher ergotropy.
          
          \par
           On the other hand, quantum renyi divergence \cite{petz1986quasi} defined by $D_{\alpha}=\frac{sgn(\alpha)}{\alpha-1}\log Tr(\rho^{\alpha}\sigma^{1-\alpha})$, is contractive under any CPTP maps for the ranges of $\alpha\in [0,2]$, i.e; $D_{\alpha}(\rho \parallel \rho_F) \geq D_{\alpha}(\Lambda(\rho) \parallel \Lambda (\rho_F))$. Eventually $D_{\alpha}$ would be a monotone under EPO i.e; $D_{\alpha}(\rho \parallel \tau^{\beta}) \geq D_{\alpha}(\Lambda(\rho) \parallel \tau^{\beta})$. For the diagonal states $D_0$ quantifies single shot extractable work under the assistance of arbitrary bath $\beta$ which decreases under $EPO$ i.e.,
       \begin{eqnarray}
       \begin{aligned}
          \rho \rightarrow \sigma~~~ \Longrightarrow~~~  & W_S(\rho) =\frac{1}{\beta}D_0(\rho \parallel \tau^{\beta}) \nonumber \\
          & \geq W_S(\sigma)= \frac{1}{\beta}D_0(\sigma \parallel \tau^{\beta}).
           \end{aligned}   
       \end{eqnarray}
       It leads to the same conclusion as given Eq.(\ref{monopas}) above for the single shot work $W_S$ and entropy when the convertibility between the two states is not possible under $EPO$. Below we provide examples for these two different situations.  \\  
       
       {\it Example1: Order of ergotropy and entropy}  
       
       Let $\rho \equiv (0.15, 0.7, 0.15)$ and  $\sigma \equiv (0.49,0.02,0.49)$ such that $\rho \nsucc \nprec \sigma$ ($\rho$ and $\sigma$ do not majorize each other), and the corresponding Hamiltonian is $H \equiv (-1,0,1)$. Then
       
       \begin{eqnarray}
       S(\rho)= 1.18129 >  S(\sigma) = 1.12144 \nonumber  \\
       W_e(\rho)= 0.55 > W_e(\sigma)= 0.47 \nonumber .
       \end{eqnarray}
       
       {\it Example2: Order of single shot work and entropy}
       
      Let $\rho \equiv (0.275,0.55,0.125,0.05)$ and $\sigma \equiv (0.35,0.35,0.3,0)$ such that $\rho \nsucc \nprec \sigma$, and $H \equiv (0,1,2,3)$. Which gives 
      
      \begin{eqnarray}
       S(\rho)= 1.57766 <  S(\sigma) = 1.58129 \nonumber  \\
       W_S(\rho) =0 < W_S(\sigma) = \frac{1}{\beta} \log (1+ \frac{e^{-2\beta}}{1+e^{-\beta}}).
      \end{eqnarray}

         \par
         Although in standard thermodynamics, the state transformations are governed by a single free energy, in finite quantum systems it is done by a series of $\alpha$-free energies \cite{horodecki2013fundamental}. Since we have considered the state transformations between $EE$ states, the role of $\alpha$-free energies is played by the $\alpha$-Renyi entropies. 
           In thermodynamics, the expansion of an ideal gas under an isothermal process produces some work due to the intake of heat from the environment. This causes the system to move towards the higher entropic states according to the second law of thermodynamics. However, in the finite quantum scenario, the system's entropy increases  under $EPO$ because although system and environment start in a product state, they become correlated though the interaction. This observation once again shows us the strong connection between informational and thermodynamical entropy.

 \section{Passive state energy as an entanglement measure}
 \subsection{Entanglement measure}
 Consider a pure state $|\psi\rangle_{AB}$ and its marginal $\rho_{A}$. 
 Define a real function $f: \mathcal{D}(\mathcal{H}_A^d) \rightarrow \mathbb{R}$, by
 \begin{equation}
 f(\rho_A)=\mathcal{E}(\psi_{AB})
 \end{equation}
 
  then $\mathcal{E}$ is said to be an entanglement measure for pure bipartite states if the function $f$ satisfies the following two conditions \cite{vidal2000entanglement},\\
  
 (i) {\it Unitary invariant:} It should be invariant under the unitary transformation of the density operator $\rho \in \mathcal{D}(\mathcal{H}^d)$ i.e., $f(\rho) = f(U\rho U^{\dagger})$.\\
 
 (ii) { \it Concavity: } It must be concave
 \begin{equation*}
 f(\sum_{i}\lambda_i \rho_i) \geq \sum_{i} \lambda_i f(\rho_i),
 \end{equation*}
 
 where $\sum_{i}\lambda_i=1~~~and~~~\lambda_i \geq 0$.

 \par  
 
  In this work we have shown (in \ref{entanglementmeasure}) that passive state energy satisfies the above criterion and acts as an entanglement monotone. This is just another entanglement monotone like entropy which is very much related with thermodynamics. Here entanglement of state $\psi_{AB}$ is defined as 
 \begin{equation}\label{pasivemonotone}
 \mathcal{E}(\psi_{AB})= E(\rho_A^p)=Tr(\rho^p_A H_A),
 \end{equation}
  where $\rho^p_A$ is the passive state corresponding to $\rho_A = Tr_B\{|\psi\rangle_{AB}\langle \psi |\}$ governed by the Hamiltonian $H_A$. For an arbitrary mixed state $\rho_{AB}=\sum_i p_i |\psi_i\rangle\langle \psi_i|$, this measure is defined by the convex hull representation \\
 \begin{equation}\label{mixedmeasure}
 \mathcal{E}(\rho_{AB})= min_{\{p_i, \psi_i\}}\sum_{i}p_i\mathcal{E}(\psi_i).
 \end{equation}
  
   \par
   
  Due to the monotonicity of passive state energy, deterministic state transformation $\psi_{AB} \rightarrow \phi_{AB}$ under LOCC always implies $E(\rho^p_A(\psi)) \geq E(\rho^p_A(\phi))$. But it has an important implication in achieving the optimal rate in the finite copy non-deterministic state transformations. If $\psi \longrightarrow \phi$ transformation is possible with optimal probability $p^{\psi \rightarrow \phi}$ then for any entanglement measure, $\mathcal{E}(\psi) \geq p^{\psi \rightarrow \phi}\mathcal{E}(\phi)$ relation must be satisfied. So the optimal probability can be defined by the minimization over all possible kinds of entanglement measures
  \begin{equation}
   p^{\psi \rightarrow \phi} \leq min_{\mathcal{E}}\{1,\frac{\mathcal{E}(\psi)}{\mathcal{E}(\phi)}\}.
  \end{equation} 
  
 For a state having $n$ number of Schmidt coefficients, one can sufficiently characterize optimal probability rate for state transformation under LOCC with $n$ number of entanglement monotones. These monotones are defined as $E_k(\psi)=\sum_{i=k}^{n}\lambda_i$, where $\lambda_i'$s are the Schmidt coefficients in decreasing order and the probability is given by,
  \begin{equation}
  p^{\psi \rightarrow \phi} = \min_{E_k}{\frac{E_k(\psi)}{E_k(\phi)}}.
  \end{equation}
  
  Passive state energies are nothing but the function of Schmidt coefficients and they matches exactly with the $E_k$'s under different system Hamiltonians. If the system governed by the Hamiltonian $H_k= \Theta^{k-1}_{i=1} \oplus \mathbb{I}^{n}_{i=k}$ then the marginal passive state energy $E(\rho^p_A)$ would be equal with the $E_k$. So, passive state energy has a direct application in finding the optimal probability for finite copy state transformation under LOCC.

  \par
  
  Passive state energy is not only a function of the spectrum but also of the energy eigenvalues. That is why its value varies with the Hamiltonian. Note that whenever we compare the entanglement between two different states, their Hamiltonians must be same. In case of comparing entanglement between states having different number $(m >n)$ of copies, extra $(|00\rangle \langle 00|)^{\otimes (m-n)}$ should be appended with the state such that marginals of both the states are governed by the same Hamiltonian, while keeping the spectrum unchanged. 
  
  \par
  
 Since the energy of a passive state is a non-additive quantity so is the corresponding entanglement measure. Suppose, $\psi_{AB}$ is a bipartite entangled state governed by the Hamiltonian $H_{AB}= H_A \otimes \mathbb{I}_B + \mathbb{I}_A \otimes H_B$. Its entanglement is quantified by the marginal's passive state energy $\mathcal{E}^1 (\psi)=E(\rho^p_A) = Tr\{\rho^p_A H_A\}$ (where the superscript $1$ denotes the number of copies), whereas for multiple copies of $\psi_{AB}$ its value would not be additive. In general for $({\psi_{AB}})^{\otimes 2}$ where the marginal state is $\rho_A \otimes \rho_A$ governed by the Hamiltonian $H_{AA}=H_A \otimes \mathbb{I}_A + \mathbb{I}_A \otimes H_A$, the entanglement of the state on an average is
  defined by
  
 \begin{equation*}
 \mathcal{E}^2(\psi)= \frac{1}{2}E({\rho^p_A}^{\otimes 2})= \frac{1}{2}Tr\{(\rho^p_A)^{\otimes 2} H_{AA}\} \leq \mathcal{E}^1(\psi).
 \end{equation*}
 
   It is a strict inequality except for the case $\mathcal{D}$$(C^2 \times C^d)$ or for some special degenerate Hamiltonian. In the asymptotic limit, passive states become the same entropic thermal states ($\tau_{\beta}, \beta \geq 0$) and entanglement is defined by the regularized version of Eq.(\ref{pasivemonotone}) as
 \begin{equation}
  \mathcal{E}^{\infty}(\psi)= E(\tau_{\beta}) = Tr(\tau_{\beta}H_A)= \lim\limits_{n \to \infty}[\frac{1}{n}Tr\{(\rho_A^p)^{\otimes n}H_{A...A}\}].
 \end{equation}

  It can be easily shown that energy of the thermal state $\tau_{\beta}$ is related with its entropy in the following way
  \begin{equation}
  E(\tau_{\beta}) = \frac{1}{\beta}\{S(\tau_{\beta})- lnZ\}.
  \end{equation}

  Since the thermal state and the marginal have same entropy then $\mathcal{E}^{\infty}(\psi)= \frac{1}{\beta}\{S(\rho_A)- lnZ\}$. 
  
  \par
  
  Finite copy state transformation is not reversible (except for the $LU$ invariant states) but in the asymptotic limit it is possible at least for the pure states. Let $\psi^{\otimes m} \rightleftarrows \phi^{\otimes n}$ by LOCC, then the amount of entanglement given by additive/non-additive measure would be same on both sides. Generally $m$ and $n$ are very large and their values are chosen such that the marginals$((\rho^{\psi}_A)^{\otimes m} / (\rho^{\phi}_A)^{\otimes n}  )$ have equivalent spectrum. If $m>n$, then to compare their entanglement we need to append $(m-n)$ number of $|00\rangle_{AB}$ states such that the spectrum of $(\rho^{\phi}_A)^{\otimes n}$ is unchanged and the Hamiltonians on both sides become same. As a consequence, their passive state energy or entanglement becomes equal i.e., $\mathcal{E}^m(\psi) = \mathcal{E}^n(\phi)$.     
  
   \subsection{Physical interpretation of our measure}
   A measure becomes more relevant when a task can be associated with it. For  example, entanglement of distillation \cite{PhysRevA.53.2046} quantifies how many Bell states one can distill from the given entangled state in the asymptotic limit by LOCC, whereas entanglement of formation\cite{PhysRevLett.80.2245} quantifies the cost of creating the desired entangled state from the Bell state. Logarithmic negativity gives the upper bound on distillable entanglement\cite{PhysRevLett.95.090503}. Relative entropy is also a useful measure of entanglement which is a monotone under all non-entangling operations\cite{brandao2008entanglement}.
   
   \par
   
   Passive state energy as an entanglement measure has an important physical interpretation in the context of thermodynamics. Suppose Alice and Bob shared a pure entangled state. They can extract work from it individually or jointly via a unitary process. Since $U_A \otimes U_B \subseteq U_{AB}$, collaboration is always useful in work extraction and the extra advantage (ergotropic gap) is coming due to the entanglement between them \cite{PhysRevA.99.052320}. Below we will show that the average ergotropic gap is equal to the passive state energy of the marginal.
   
   \par
   
   For maximal work extraction from a closed system via a unitary, the system should reach to the same spectral but minimum energetic state (passive state). Maximum extractable work obtained locally by Alice and Bob is defined by 
   \begin{eqnarray}\label{local ergotropy}
   \begin{aligned}
   & W_e(A)+W_e(B) \nonumber \\
   & = Tr(\rho_A H_A)-Tr(\rho^p_A H_A)
    +Tr(\rho_B H_B)-Tr(\rho^p_B H_B)
    \end{aligned}
   \end{eqnarray}

   where $\rho_A=Tr_B(|\psi\rangle_{AB} \langle \psi|)$ and $\rho_B=Tr_A(|\psi\rangle_{AB} \langle \psi|)$ are the marginals of Alice and Bob respectively while $|\psi \rangle_{AB}$ is the pure entangled state shared between them.
   
   \par
   
  From $|\psi\rangle_{AB}$ one can reach $|00\rangle_{AB}$ to extract maximum global work 
   \begin{equation}\label{global ergo}
    W_e(AB)= Tr(|\psi\rangle_{AB} \langle \psi| H_{AB})
   \end{equation}
   where $|00\rangle_{AB}$ is the zero energetic eigenstate of $H_{AB}$.
   So the ergotropic gap in general is defined as 
   \begin{equation}\label{ergotropicgap}
   \Delta^{EG}(\psi_{AB})= Tr(\rho^p_A H_A)+ Tr(\rho^p_B H_B).
   \end{equation}
   
   Since the shared state is pure entangled, the marginals have same spectrum and if their Hamiltonians are given by $H_A=\sum\limits_{i=1}^{d_A-1}\epsilon_{i}|\epsilon_{i}\rangle\langle \epsilon_{i}|$ and $H_B=\sum\limits_{i=1}^{d_B-1}\epsilon_{i}|\epsilon_{i}\rangle\langle \epsilon_{i}|$, then their passive state energy should be equal i.e.
   
   \begin{equation}
   \Delta^{EG}= 2Tr(\rho^p_A H_A)=2E(\rho_A^p).
   \end{equation} 
   
   Since the passive state energy
    quantifies the average work gain in collaboration, we would call it a {\it thermodynamic measure of entanglement}. Alternatively, ergotropic gap can be considered as an entanglement measure since it is proportional to the passive state energy. For the mixed state thermodynamic measure of entanglement is given by Eq.[\ref{mixedmeasure}]. However unlike the pure case it is not equivalent to the ergotropic gap or passive state energy of the given mixed state. Their exist classically correlated states whose ergotropic gap is non zero. 
    
     \subsection{Faithfulness and Monogamous}
     
     An entanglement measure is called faithful if and only if its value becomes zero for all separable states, i.e.,
     $$\mathcal{E}(\rho_{AB})=0 ~~~iff~~~\rho_{AB}\in SEP.$$ 
     \par
     Entanglement of formation and negativity are example of this kind. Monogamy is a fundamental property of quantum states just like no cloning, superposition etc. In the classical regime, party $A$ can be maximally correlated with parties $B$ and $C$ simultaneously, which is not allowed in the quantum domain due to its inherent mathematical structure. If two parties share a 
     pure entangled state then any third party would be completely uncorrelated with them. Generally if we consider a three-party entangled state $\rho_{ABC}$, then the sum of entanglements  $A$ vs. B and A vs. C is always upper bounded by the entanglement of A vs. BC. If an entanglement measure shows this trait then it is said to be monogamous
     \begin{equation}\label{generalmonogamous}
     \mathcal{E}(\rho_{A|BC}) \geq \mathcal{E}(\rho_{A|B})+\mathcal{E}(\rho_{A|C}),
     \end{equation}
     
     where $\mathcal{E}(\rho_{X|Y})$ quantifies the entanglement of $X$ vs. $Y$.
    For example entanglement of distillation and squashed entanglement \cite{brandao2011faithful} are monogamous measures.  
     
     \par
     
     Winter et.al. have  defined monogamy in a more general way and shown that a measure cannot be faithful and monogamous simultaneously \cite{PhysRevLett.117.060501}. An entanglement measure $\mathcal{E}$ is called monogamous if there exists a nontrivial function $f:\mathbb{R}_{\geq 0}\times \mathbb{R}_{\geq 0} \rightarrow \mathbb{R}_{\geq 0}$ such that the generalized monogamy relation 
     \begin{equation}\label{winter}
      \mathcal{E}(\rho_{A|BC}) \geq f(\mathcal{E}(\rho_{A|B}),\mathcal{E}(\rho_{A|C})),
     \end{equation}
    is satisfied for any state $\rho_{ABC}$ $\in$ $ \mathcal{D}(\mathcal{H_A} \otimes \mathcal{H_B} \otimes \mathcal{H_C})$. According to this definition, not only our thermodynamic measure, but also many other important measures like Concurrence loose their generalized monogamy (Eq.\ref{winter}) since they are faithful. It is however well-known that concurrence does not follow Eq.(\ref{generalmonogamous}) but its square follows this for $2 \times 2 \times 2$ systems \cite{PhysRevA.61.052306}. A physically motivated but weaker monogamous criterion has been given in \cite{PhysRevA.91.012339,gour2018monogamy,PhysRevA.99.042305} which states that an entanglement measure $\mathcal{E}$ is monogamous if for any $\rho_{ABC} \in  \mathcal{D}(\mathcal{H_A} \otimes \mathcal{H_B} \otimes \mathcal{H_C})$ that satisfies, 
    \begin{equation}\label{disentangling}
    \mathcal{E}(\rho_{A|BC})=\mathcal{E}(\rho_{A|B})
    \end{equation}  
      then $\mathcal{E}(\rho_{A|C})=0$.
      \par
      The above definition led to a more quantitative relation which is similar to Eq. (\ref{generalmonogamous}). Let $\mathcal{E}$ be a continuous measure of entanglement. Then, $\mathcal{E}$ is monogamous according to Eq. (\ref{disentangling}) if and only if there exist $0<\gamma < \infty$ such that 
     \begin{equation}\label{monogamous}
      \mathcal{E}^{\gamma}(\rho_{A|BC}) \geq \mathcal{E}^{\gamma}(\rho_{A|B})+\mathcal{E}^{\gamma}(\rho_{A|C}).
      \end{equation}
     for all $\rho_{ABC} \in \mathcal{D}(\mathcal{H_A} \otimes \mathcal{H_B} \otimes \mathcal{H_C})$ with fixed dim $\mathcal{H_{ABC}}=d < \infty$.
     Although every entanglement monotone does not satisfy Eq.(\ref{winter}) but for fixed dimension there always exists a $\gamma$ for which Eq.(\ref{monogamous}) is satisfied. Our entanglement measure shows the second kind of monogamy and there exists an $\gamma$ for which ergotropic gap should follow Eq.(\ref{monogamous}) for a certain dimension.
     
     \par
     
      For mixed states our thermodynamic measure is represented by convex hull given by Eq.(\ref{mixedmeasure}) where minimization is over all pure state decompositions of the given state. Finding analytical solution is not easy even for lower dimension. But one can obtain some numerical estimate for the value of $\gamma$ by semi definite programming. However since this is not the focus of the present paper, we leave it as an open problem for the future. But for curiosity we investigate what kind of relation exists for arbitrary dimensional multipartite systems if we take $\gamma=1$.

      Let $|\psi\rangle_{ABC}$ be a three qubit entangled state where each qubit is governed by the Hamiltonian $H_i=|1\rangle \langle 1|$. Then its entanglement in any bi-partite cut is given by
      \begin{equation}\label{ergomono}
      \Delta_{A|BC}^{EG}(\psi)= \Delta_{A|B}^{EG}(\rho_{AB}) + \Delta_{A|C}^{EG}(\rho_{AC}).
      \end{equation}
     Note that on the R.H.S. $\Delta_{X|Y}^{EG}(\rho_{XY})$ is not an entanglement measure but rather just the ergotropic gap of system $\rho_{XY}$. It also tells us that for fixed entanglement between $A|BC$, there always exists a trade off in ergotropic gap for the system $AB$ and $AC$. In the extreme case, if the joint state for any pair (e.g. $AC$) is passive then the presence of $C$ is not necessary to get the collaborative advantage. For the GHZ class this feature can never be seen since all bipartite marginals have non zero ergotropic gap. But there exists W class for which only one of the bipartite marginals can be passive. For example the state
     \begin{equation*}
     |\psi\rangle_{ABC}= \sqrt{\lambda_1}|001\rangle +\sqrt{\lambda_2}|010\rangle +\sqrt{\lambda_3}|100\rangle
     \end{equation*}
     has 
     bipartite marginals
     \begin{eqnarray}
     \rho_{AB}=\lambda_1 |00\rangle \langle 00|+(\lambda_2+\lambda_3)|\phi\rangle \langle \phi| \nonumber \\
     \rho_{AC}=\lambda_2 |00\rangle \langle 00|+(\lambda_1+\lambda_3)|\eta\rangle \langle \eta| \nonumber \\
     \rho_{BC}=\lambda_3 |00\rangle \langle 00|+(\lambda_1+\lambda_2)|\chi\rangle \langle \chi|
     \end{eqnarray}
     where 
     
     \begin{eqnarray}
     |\phi\rangle = \sqrt{\frac{\lambda_2}{\lambda_2+\lambda_3}}|01\rangle + \sqrt{\frac{\lambda_3}{\lambda_2+\lambda_3}}|10\rangle
     \nonumber \\
      |\eta\rangle =\sqrt{\frac{\lambda_1}{\lambda_1+\lambda_3}}|01\rangle + \sqrt{\frac{\lambda_3}{\lambda_1+\lambda_3}}|10\rangle 
      \nonumber \\
        |\chi\rangle =\sqrt{\frac{\lambda_1}{\lambda_1+\lambda_2}}|01\rangle + \sqrt{\frac{\lambda_2}{\lambda_1+\lambda_2}}|10\rangle. 
     \end{eqnarray}
     Here if any one of the $\lambda_i$ is greater or equal to $\frac{1}{2}$ then there would exist one passive state and the corresponding ergotropic gap would be zero. Suppose $\lambda_2 \geq \frac{1}{2}$ then $\Delta_{A|C}^{EG}(\rho_{AC})=0$ and according to Eq.(\ref{ergomono}) entanglement between $A|BC$ would be equal to the ergotropic gap of AB i.e., $\Delta^{A|BC}_{EG}(\psi)= \Delta_{EG}(\rho^{AB})$. We have discussed above a particularly simple system but Eq.(\ref{ergomono}) can be generalized for arbitrary dimensional tripartite pure systems, where it can be easily proved that entanglement between any bipartite cut is upper bounded by the sum of ergotropic gap of those individuals as shown in the \ref{tripartite}.

     \subsection{Three qubit state classification}
      We know that for the three qubit pure state, there exist six classes (under $SLOCC$) among which two are genuinely entangled, three are bi-separable and one is product\cite{PhysRevA.62.062314}. Except the genuine classes, distinction among all the rest can be made via local entropies. These two classes namely GHZ and W can be distinguished by a measure called {\it tangle} which gives non zero value only for the former \cite{PhysRevA.61.052306}. We shall approach this classification from the thermodynamic perspective. Since the ergotropic gap is a good measure for bipartite pure entangled states, it can shed some light on the above problem. We have considered the GHZ, W and bi-separable states of the following kind
      \begin{eqnarray}\label{class}
      \begin{aligned}
       |\psi\rangle_{GHZ} = & \sqrt{\lambda_{max}} |\psi_1\rangle |\psi_2\rangle |\psi_3\rangle \nonumber \\
      &+ e^{i\phi} \sqrt{\lambda_{min}} |\psi_1\rangle^{\perp} |\psi_2\rangle^{\perp} |\psi_3\rangle^{\perp} \nonumber \\
       |\psi\rangle_{W} = & \sqrt{\lambda_1}|\psi_1 \rangle |\psi_2 \rangle |\psi_3 \rangle^{\perp}  
       +e^{i\phi_1}\sqrt{\lambda_2}|\psi_1 \rangle |\psi_2 \rangle^{\perp} |\psi_3 \rangle \nonumber \\
       & +e^{i\phi_2}\sqrt{\lambda_3}|\psi_1 \rangle^{\perp} |\psi_2 \rangle |\psi_3 \rangle \nonumber \\
       |\psi\rangle_{AB} \otimes |\phi\rangle_C = & (\sqrt{p_{min}}|\psi_1\rangle|\phi_1\rangle + \sqrt{p_{max}}|\psi_1\rangle^{\perp}|\phi_1\rangle^{\perp})_{AB}\nonumber \\
       & \otimes |\phi\rangle_C
         \end{aligned}
      \end{eqnarray}
      where $\lambda_1 \geq \lambda_2 \geq \lambda_3$.
      
      In the table below, we make the distinction among various classes using ergotropic gap between one party vs. the rest $(\Delta_{X|YZ}^{EG})$, calculated via Eq.(\ref{ergotropicgap}).\\

    \begin{widetext}
     \begin{tabular}{|p{5cm}||p{3cm}|p{3cm}|p{3cm}|}
     \hline
      Class& $\Delta_{A|BC}^{EG}$ & $\Delta_{B|AC}^{EG}$ & $\Delta_{C|AB}^{EG}$\\
      \hline
      $|\psi\rangle_{GHZ}$  & $2\lambda_{min}$    & $2\lambda_{min}$ & $2\lambda_{min}$\\
      \hline
      $|\psi\rangle_{W}(\lambda_1 \geq \frac{1}{2})$ & $2\lambda_{3}$   & $2\lambda_{2}$  & $2(\lambda_{2}+\lambda_3)$\\
       $|\psi\rangle_{W}(\lambda_1 \leq \frac{1}{2})$ & $2\lambda_{3}$   & $2\lambda_{2}$  & $2\lambda_{1}$\\
      \hline
     $|\psi\rangle_{AB} \otimes |\phi\rangle_C$   & $2p_{min}$ & $2p_{min}$ &  0\\
     \hline
      $|\psi\rangle_{AC} \otimes |\phi\rangle_B$ &  $2p_{min}$   & 0 & $2p_{min}$ \\
     \hline
     $ |\phi\rangle_A \otimes |\psi\rangle_{BC}$& 0  & $2p_{min}$  & $2p_{min}$ \\
      \hline
      $|\psi\rangle_A \otimes |\psi\rangle_B \otimes |\psi\rangle_C$ & 0 & 0& 0 \\
      \hline
     \end{tabular} \label{table}
     \end{widetext} 
     Note that as expected, our thermodynamic measure gives equal value for all bipartite cuts of the given GHZ class since they have equal entanglement. On the other hand, for the W class (except $\lambda_1=\lambda_2=\lambda_3=\frac{1}{3}$) all bipartite cuts give different values. For this state it gives the same value as the GHZ state $|\psi\rangle_{GHZ} = \sqrt{\frac{2}{3}} |\psi_1\rangle |\psi_2\rangle |\psi_3\rangle + \sqrt{\frac{1}{3}} |\psi_1\rangle^{\perp} |\psi_2\rangle^{\perp} |\psi_3\rangle^{\perp}$ and hence they cannot be distinguished. One way to facilitate the distinction is to first dephase them in their own basis and then calculate the extractable work difference between the global and individual local states. The dephased W and GHZ states are of the following form
      \begin{eqnarray}
      \begin{aligned}
     \rho_{w}=& \frac{1}{3}|\psi_1\psi_2\psi_3^{\perp}\rangle \langle \psi_1\psi_2\psi_3^{\perp}| \nonumber \\
     & +
     \frac{1}{3}|\psi_1\psi_2^{\perp}\psi_3 \rangle \langle \psi_1\psi_2^{\perp}\psi_3| + \frac{1}{3}|\psi_1^{\perp}\psi_2\psi_3 \rangle \langle \psi_1^{\perp}\psi_2\psi_3 | \nonumber \\
     \rho_{ghz}
      = & \frac{2}{3}|\psi_1 \psi_2 \psi_3 \rangle \langle \psi_1 \psi_2 \psi_3|+ \frac{1}{3} |\psi_1^{\perp} \psi_2^{\perp} \psi_3^{\perp} \rangle \langle \psi_1^{\perp} \psi_2^{\perp} \psi_3^{\perp} | .
      \end{aligned}
      \end{eqnarray} 
      
      And the work difference is defined by 
      \begin{equation*}
      \Delta_{A|B|C}^{EG}= Tr(\rho^A_pH_A) + Tr(\rho_p^BH_B) + Tr(\rho_p^CH_C)-Tr(\rho_p H_{ABC}).
      \end{equation*} 
      
      For the dephased W state this value is $\Delta_{A|B|C}^{EG}(\rho_w)=\frac{1}{3}$ where as for the dephased GHZ state, $\Delta_{A|B|C}^{EG}(\rho_{ghz})=\frac{2}{3}$. 
      \par
     The above analysis can be extended to distinguish the GHZ and Dicke class of states (generalization of Eq.(\ref{class})) in the multipartite setting.

 \section{Conclusion}
 
 In this article, it has been elaborately shown that for the finite $ EE $ quantum states, ordering between them on the basis of extractable work is completely independent with the order on entropy. To illustrate this, a resource theoretic framework is employed, where $EPO$ has been taken as the free operation. If a state is convertible to another state by the free operation, then entropy and extractable work become equivalent like in the asymptotic limit. Here majorization naturally turns out to be a sufficient criterion for the convertibility (one way) between two states under $EPO$. Eventually entropy, ergotropy, single shot work under bath all become independent monotones of $EPO$. This also means that violation of this criterion make two states incomparable under $EPO$, thereby rendering the ordering of work and entropy completely independent. When a system moves under $EPO$, it's entropy (work content) eventually increases (decreases) because of the correlations that build up with the external environment. The additional constraint of conservation of system energy makes our resource theory a special case of resource theory of athermality. If we consider degenerate Hamiltonian, our resource theory leads to the resource theory of purity. That's why just like the $\alpha$ free energies, here Renyi entropies characterize infinite second laws.
 \par
 In the second part, we have shown that passive state energy is a good entanglement measure for the pure bi-partite states since it is a concave function and invariant under unitary. It is non additive, a feature that makes it possible to generate Vidal's monotones which characterizes the optimal rate on entangled state transformation under LOCC. The passive state energy characterizes the ergotropic gap for the pure entangled states which makes it a thermodynamic measure of entanglement independent of entropy. Since it is proportional with the ergotropic gap, one can give it the same status. We further generalized this measure for the bipartite mixed states via the convex hull extension. Due to faithfulness, it loses it's monogamous nature and for what value of $\gamma$ it shows dimension dependent monogamy is left as an open problem. For the special case $\gamma=1$, the entanglement between any cut is bounded by the sum of the ergotropic gap of corresponding individuals for pure tripartite states. We have further shown that just like entropy, the ergotropic gap can be useful to distinguish the three qubit pure entangled states of special kinds.
   
 \section{Acknowledgement}
 M.A. would like to thank Swapan Rana and Anandamay Das Bhowmik for fruitful discussions and acknowledges financial support from the CSIR project 09/093(0170)/2016-EMR-I.

  \section*{References}
  	%
  	

\begin{thebibliography}{43}
\bibitem [{ci()}]{ci}%
  		\BibitemOpen
  		\bibfield  {title} {\enquote {\bibinfo {title} {Complete informtaion means
  					that exact values of the system energy as well as the corresponding passive
  					state energy of $\rho$ and $\sigma$ is required.}}\ }\href@noop {} {\
  		}\BibitemShut {NoStop}%
\end{thebibliography}
 
 \onecolumngrid
 \appendix
  
  \section{Free states}
  
  \subsubsection*{Definition of a free state:} 
  Under a free operation states $\rho_F$ would be called free if their presence does not increase the resource $R$ of a given system $\rho_S$ and the operation on free state remains free i.e.,\\
  (i) $\Lambda(\rho_S)=\sigma_S= Tr_F\{U_{SF}(\rho_S \otimes \rho_F)U^{\dagger}_{SF}\} \Longrightarrow R(\rho_S) \geq R(\sigma_S).$\\
  (ii) $\Lambda(\rho_F) = \sigma_F$
  
  \subsubsection*{Theorem: The states  $\tau_E^{\beta}=e^{-\beta H_E}/tr(e^{-\beta H_E})$ are the free states under the free operation EPO, where $\beta \in [-\infty,+\infty]$ .}\label{theorem}

  \par

  {\it Proof:} We have considered $EPO$ is a free operation which itself is a restricted thermal operation. Since it is an unital map and sufficiently characterized by the majorization criterion, entropy would increase as the system evolves. So the free states would have highest entropy. In this resource theory, we have considered system and ancilla (free states) in uncorrelated state and forms a closed system which evolves under the energy preserving unitary such that
     \begin{eqnarray*}
     S(\rho_S \otimes \rho_F)  = S\{U(\rho_S \otimes \rho_F)U^{\dagger}\}
     \end{eqnarray*}  
     
     Since after evolution system and free state get correlated they possess non zero mutual information which implies,
     
     \begin{equation*}
     S(\rho_S) + S(\rho_F)  \leq  S(\sigma_S) + S(\rho'_F)
     \end{equation*}
     
     Since we have considered cyclic process we want at least same purity back for the ancillary system after operation i.e., $S(\rho_{F})=S(\rho'_{F})$, which leads to
     \begin{equation}\label{rhosig}
     S(\rho_S) \leq S(\sigma_S).
     \end{equation}

     Now if we consider $\rho_F$ instead of $\rho_S$ in Eq.(\ref{rhosig}), then $\sigma_S = \rho''_F$ because this are the maximum entropic states. But we know that for constant energy the state $\tau_E^{\beta}=e^{-\beta H_E}/tr(e^{-\beta H_E})$ has the maximum entropy. Positive $\beta$'s are the thermal state and considered as free in resource theory of athermal states whereas in this resource theory we take $-\infty \leq \beta \leq \infty$ since all the $\beta$ has different energy. 
     
     \section{Passive state energy is a monotonic function under EPO} \label{passivemonotone}
     
     We need to show that passive state energy of a given system increases as it evolves under $EPO$. Since the state transformation under $EPO$ is necessarily characterized by the majorization criterion, we need to prove that
     
     \begin{equation}
    \rho \succ \sigma \Longrightarrow E(\rho^p) \leq E(\sigma^p)
     \end{equation}\label{pmono}
     and this suffices to demand that, passive state energy is monotone under any unital map, i.e., under EPO.
     \par
     Let state $\rho \equiv (p_1,p_2, \cdots , p_n)$ and $\sigma \equiv (q_1,q_2, \cdots , q_n)$ have spectrum in non-increasing order and extra zeros are appended to make the dimensions equal. They are governed by the same marginal Hamiltonian $H= \sum\limits_{i}\epsilon_{i}|i\rangle \langle i|$, $\epsilon_i \leq \epsilon_{i+1}$. The corresponding passive energy is $E(\rho^p)= \sum_i p_i \epsilon_i$ and $E(\sigma^p)= \sum_i q_i \epsilon_i$, and their difference is 
       \begin{eqnarray}
       \begin{aligned}
   E(\sigma^p)-E(\rho^p) & = \sum_i q_i \epsilon_i - \sum_i p_i \epsilon_i \nonumber \\
      & = - \sum_i(p_i-q_i)\epsilon_i \nonumber \\
      & = - (p_1-q_1)\epsilon_1 + (p_1-q_1)\epsilon_2 -(p_1+p_2-q_1-q_2)\epsilon_2 + (p_1+p_2-q_1-q_2)\epsilon_3 \nonumber \\
       & +\cdots -\sum\limits_{i=1}^{k<n-1}(p_i-q_i) \epsilon_k + \sum\limits_{i=1}^{k<n-1}(p_i-q_i) \epsilon_{k+1} + \cdots - \sum\limits_{i=1}^{n-1}(p_i-q_i) \epsilon_{n-1} - (p_n-q_n) \epsilon_{n} \nonumber \\
       & =\sum\limits_{j=1}^{n-1}\sum\limits_{i=1}^{j} (p_i-q_i)(\epsilon_{j+1}-\epsilon_{j})
        \end{aligned}
       \end{eqnarray}
     
         If $\sum\limits_{i=1}^{j} (p_i-q_i) \geq 0$, $\forall j \in [1,n-1]$, which is nothing but the $\rho \succ \sigma $ criterion, then $E(\rho^p)\leq E(\sigma^p)$.

       \section{Passive state}
       {\bf{\it Schur-Horn theorem:}} Let ${\displaystyle {d} =\{d_{i}\}_{i=1}^{N}}$ and ${\lambda}=\{\lambda_i\}_{i=1}^N$ be vectors in $\mathbb{R}^{N}$ such that their entries are in non-increasing order. There is a Hermitian matrix with diagonal values $\{d_i\}_{i=1}^N$ and eigenvalues $\{\lambda_i\}_{i=1}^N$ if and only if 
       
            $${\displaystyle \sum _{i=1}^{n}d_{i}\leq \sum _{i=1}^{n}\lambda _{i}\qquad n=1,2,\ldots ,N}$$
       and
       
       $${\displaystyle \sum _{i=1}^{N}d_{i} = \sum _{i=1}^{N}\lambda _{i}\qquad}.$$
       
       \subsection{Passive state is the lowest energetic state which is diagonal in energy basis}
       Work extraction from a closed system $\rho$ by a unitary operation does not change its spectrum. In the maximum work extraction, the system moves to the lowest energetic state which is called passive state defined by $\rho^p$. With the help of Schur-Horn theorem it can be shown that the passive state is diagonal in energy basis.\\
       
        Let $\rho$ has the spectrum $\lambda \equiv \{\lambda_i\}_{i=1}^N$(arranged in non-increasing order), transform to the arbitrary state $\sigma=U\rho U^{\dagger}$ having diagonal element $q \equiv \{q_{i}\}_{i=1}^{N}$ (arranged in the non-increasing order) . Passive state $\rho^p = \sum\limits_{i=1}^{N}  \lambda_i |i\rangle \langle i|$ and the corresponding system Hamiltonian $H_S = \sum\limits_{i=1}^{N}  \epsilon_i |i\rangle \langle i|$ where, $\epsilon_i \leq \epsilon_{i+1}.$  According to Schur-Horn theorem $q \prec \lambda$ whereas using  Eq.[\ref{pmono}] we can say that $Tr(\sigma H_S) \geq \sum\limits_{i=1}^{N}q_i \epsilon_i \geq \sum\limits_{i=1}^{N}  \lambda_i \epsilon_i$. It proves that passive state, is the lowest energetic state diagonal in energy basis.
        
       \subsection{Passive state energy is an entanglement measure}\label{entanglementmeasure}
       
     {\it Definition:} Let us define a real function $f: \mathcal{D}(\mathcal{H}_A^d) \rightarrow \mathbb{R}$, by
     \begin{equation*}
     f(\rho_A)=\mathcal{E}(\psi_{AB}).
     \end{equation*}
     
      Then $\mathcal{E}$ is an entanglement measure if $f$ satisfies \cite{vidal2000entanglement} ;\\
     (i) {\it Unitary invariant:} $f(\rho) = f(U\rho U^{\dagger})$.\\
     (ii) { \it Concavity: }$
     f(\sum_{i}\lambda_i \rho_i) \geq \sum_{i} \lambda_i f(\rho_i)$, where $\sum_{i}\lambda_i=1$ and $\lambda_i \geq 0$.
      \par
      
      Since the spectrum remains invariant under the unitary action, the respective passive state an its energy would be unique.
      
      Here we first define Vidal's entanglement monotones $\{E_k\}^n_{k=1}=\sum\limits_{i=k}^{n} p_i$ where $p_i$'s are the Schmidt number of bi-partite state $|\psi\rangle_{AB} \in \mathbb{C}^n\times \mathbb{C}^m$ $(n \leq m)$ \cite{PhysRevLett.83.1046}. Consider that the state $\rho_A \equiv (p_1,p_2, \cdots , p_n)$, which is a marginal of $|\psi_{AB}\rangle $, majorizes $\sigma_A \equiv (q_1,q_2, \cdots , q_n)$, marginal of $|\phi_{AB} \rangle $ i.e., $(\rho_A \succ \sigma_A)$ which implies $E_k(\rho_A) \leq E_k(\sigma_A)$ for all $k \in [1,n]$. It follows that $E_k$'s are the Schur concave functions i.e.,
      
      \begin{eqnarray}\label{pe}
      \begin{aligned}
      \sum\limits_{i=k}^{n}t_i & \geq \lambda \sum\limits_{i=k}^{n}r_i+ (1-\lambda)\sum\limits_{i=k}^{n}s_i=\sum\limits_{i=k}^{n}x_i \nonumber \\
      & \Rightarrow
      \sum\limits_{i=1}^{k-1}t_i \leq \sum\limits_{i=1}^{k-1}x_i \nonumber \\
       & \Rightarrow
      t \prec x
      \end{aligned}
      \end{eqnarray}

      where $\tau = \lambda \tau_1 + (1-\lambda) \tau_2$, $\tau = \{t_i\}^n_{i=1}$, $\tau_1=\{r_i\}^n_{i=1}$ and $\tau_2=\{s_i\}^n_{i=1}$.  
      
      \par
      From Eq.[\ref{pmono}] and Eq.[\ref{pe}],
      \begin{eqnarray}
      \begin{aligned}
       & \sum\limits_{i=1}^{n}t_i\epsilon_i \geq \sum\limits_{i=1}^{n}x_i\epsilon_i \nonumber \\
       \Rightarrow & \sum\limits_{i=1}^{n}t_i\epsilon_i \geq \lambda \sum\limits_{i=1}^{n}r_i\epsilon_i+ (1-\lambda)  \sum\limits_{i=1}^{n}s_i\epsilon_i \nonumber \\
       \Rightarrow & E_p(\tau) \geq \lambda E_p(\tau_1) + (1-\lambda) E_p(\tau_2)
       \end{aligned}
      \end{eqnarray}
      It shows that passive state energy is a concave function and hence acts as an entanglement measure. The above concavity proof also follows straight forwardly from \ref{passivemonotone} but here we tried to show a connection between passive state energy and Vidal's monotones $E_k$'s.
      
      \section{$\Delta_{A|BC}^{EG}(\psi) \leq \Delta_{A|B}^{EG}(\rho^{AB}) + \Delta_{A|C}^{EG}(\rho^{AC})$}\label{tripartite}
      
      For a tripartite pure state $|\psi \rangle_{ABC}$, spectrum of any bipartite cut is always equal i.e., $\lambda(\rho_A)=\lambda(\rho_{BC})$. If we consider the equal marginal Hamiltonian $H= \sum_{i}i|i\rangle \langle i |$ for every party, then their corresponding passive state energy would follow the order
       \begin{equation}
      E(\rho^p_A) \geq E(\rho^p_{BC})
      \end{equation}
      where the equality would hold for the $2 \times 2 \times 2$ systems only.
      
      Ergotropic gap of a bi-partite system $\rho_{XY}$ is given by
      \begin{eqnarray}
      \begin{aligned}
      \Delta_{X|Y}^{EG}(\rho_{XY}) & = W_e(\rho_{XY})-W_e(\rho_{X})-W_e(\rho_{Y}) \nonumber \\
      & = E(\rho^p_{X})+E(\rho^p_{Y})-E(\rho^p_{XY})
      \end{aligned}
      \end{eqnarray}
      where $E(\rho^p_{XY})=0$ if $\rho_{XY}$ becomes pure.\\

      For pure tripartite systems, according to Eq.(\ref{ergotropicgap}), entanglement between $A|BC$ is given by $\Delta^{A|BC}_{EG}(\psi)= Tr(\rho^p_A H_A)+Tr(\rho^p_B H_B)$ which is not true for the mixed state. Now,
      \begin{eqnarray}
      \begin{aligned}
      \Delta_{A|B}^{EG}(\rho_{AB}) + \Delta_{A|C}^{EG}(\rho_{AC})-\Delta_{A|BC}^{EG}(\psi_{ABC}) & = E(\rho^p_{A})+E(\rho^p_{B})-E(\rho_{AB}^p)
      +E(\rho_{A}^p)+E(\rho_{C}^p)-E(\rho_{AC}^p)
      -E(\rho_{A}^p)-E(\rho_{BC}^p) \nonumber \\
    & = \{E(\rho_{A}^p)-E(\rho_{BC}^p)\}+\{E(\rho_{B}^p)-E(\rho_{AC}^p)\}+\{E(\rho_{C}^p)-E(\rho_{AB}^p)\}  \geq 0     \nonumber \\
     & \Rightarrow
     \Delta_{A|BC}^{EG}(\psi_{ABC}) \leq \Delta_{A|B}^{EG}(\rho_{AB}) + \Delta_{A|C}^{EG}(\rho_{AC})
     \end{aligned}
      \end{eqnarray}

\end{document}